\documentclass[
    ,final            
    ,numbers              
  ]
  {aipproc}

\usepackage{amssymb,amsmath,amstext,amsthm,amsfonts}
\layoutstyle{6x9}

\usepackage[usenames]{color}
\definecolor{magenta}{cmyk}{0,1,0,0}

\newcommand{\plotscale}{.60}
\newcommand{\atom}[2]{\mbox{$^{#1}\text{#2}$}}
\newcommand{\carbon}{{\atom{12}{C}}}
\newcommand{\oxygen}{{\atom{16}{O}}}

\newcommand{\reffig}[1]{Fig.~\ref{#1}}

\newcommand{\refcite}[1]{Ref.~\cite{#1}}

\begin{document}

\title{Entanglement of Quasielastic Scattering\\ and Pion Production}


\author{Ulrich Mosel}{
  address={Institut fuer Theoretische Physik, Universitaet Giessen, D-35392 Giessen, Germany}
}

\author{Olga Lalakulich}{
  address={Institut fuer Theoretische Physik, Universitaet Giessen, D-35392 Giessen, Germany}
}

\author{Tina Leitner}{
  address={Institut fuer Theoretische Physik, Universitaet Giessen, D-35392 Giessen, Germany}
}

\begin{abstract}
The extraction of neutrino oscillation parameters requires the determination of the neutrino energy from observations of the hadronic final state. Here we discuss the difficulties connected with this energy reconstruction for the ongoing experiments MiniBooNE and T2K. We point out that a lower limit to the uncertainty in the reconstructed energy from Fermi motion alone amounts to about 15\%. The entanglement of very different elementary processes, in this case quasielastic scattering and pion production, in the actual observables leads to considerably larger errors. We discuss the sensitivity of the energy reconstruction to detection techniques and experimental acceptances. We also calculate the misidentification cross section for electron appearance in the T2K experiment due to neutral pion production.
\end{abstract}

\maketitle


\section{Introduction}

Long-baseline experiments with neutrino beams use -- due to the neutrino production mechanism -- necessarily broad-band beams. This means that the neutrino (and antineutrino) energy is not sharp, but instead distributed over a fairly wide energy range. For example, the MiniBooNE experiment at Fermilab has an average neutrino energy of about 700 MeV, but the energy distribution extends all the way up to 2 GeV and beyond. This poses a challenge for all experiments that try to determine neutrino oscillation parameters from such experiments since the neutrino beam energy enters as a crucial parameter into all oscillation formulas. The only way out is then to try to reconstruct the neutrino energy from the hadronic and leptonic final state by assuming one special reaction type and using quasifree kinematics for this reaction on a target nucleon at rest. However, since all presently running experiments use nuclear targets the target nucleons all have momenta up to the Fermi momentum ($\approx 250$ MeV) so that there is a natural smearing of reconstructed energies even if the incoming energy distribution were sharp. In addition, the initial, primary final state is not observable. The same holds for the actual reaction mechanism of the first interaction of the incoming neutrino with a target nucleon. What is observable are knock-out nucleons and mesons that are created either in the first interaction (with a priori unknown reaction mechanism) or in some final state interaction (fsi). Various reaction mechanisms, such as quasielastic scattering (QE) or pion production dominate in different regimes of the energy transfer so that any broad-band experiment necessarily contains contributions from different reaction mechanisms. To isolate quasielastic scattering, for example, for use in the neutrino energy reconstruction then requires the use of event generators which are reliable not only for one particular reaction type, but are well tested and reliable for a broad class of relevant reactions. In particular, the errors connected with the use of these generators have to be well under control \cite{Harris:2004iq,Tanaka:2009zzb,FernandezMartinez:2010dm}.

\section{GiBUU transport method}

To account for all the different reaction types we use the GiBUU transport
model \cite{Buss:2011mx}, where the neutrino first interacts with one bound nucleon at a time (impulse approximation (IA)).  The use of the IA requires a good description of both
the elementary vertex and the in-medium-modifications. The final state of this initial
reaction undergoes complex hadronic final-state interactions.

The GiBUU model is based on well-founded theoretical ingredients and has been tested in
various very different nuclear reactions; in particular, against electron- and
photon-scattering data \cite{Buss:2011mx,Leitner:2008ue,Krusche:2004uw}. The GiBUU model has been shown to work
very well for QE scattering of electrons if the momentum-transfer is larger than about 300 MeV; comparison with
photoproduction experiments for $\pi^0$ on nuclear targets shows that the cross section again is described very well up
to the $\Delta$ resonance peak, but comes out too low on the high-energy side of the $\Delta$ by up to 20\% for light nuclei \cite{Krusche:2004uw}. These shortcomings may be due to a breakdown of the impulse approximation \cite{Ankowski:2010yh}. For more details on GiBUU we refer the reader to \refcite{Buss:2011mx}.

\section{QE-pion Entanglement and Energy Reconstruction}
In this section we now investigate how clean the experimental QE-like events actually are, to what degree they depend on the event reconstruction through some generator and how big the inaccuracies in the energy reconstruction actually are.
We follow here closely the presentation in \cite{Leitner:2010kp}.

\begin{figure}[tbp]
  \centering
  \includegraphics[scale=\plotscale]{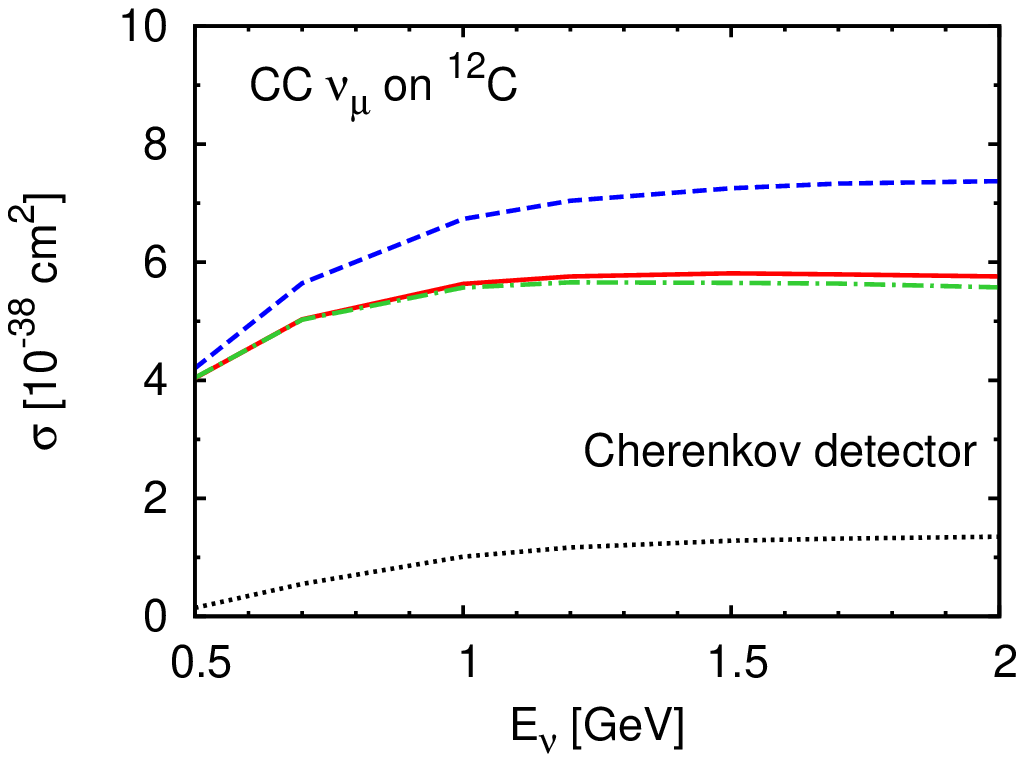}
  \includegraphics[scale=\plotscale]{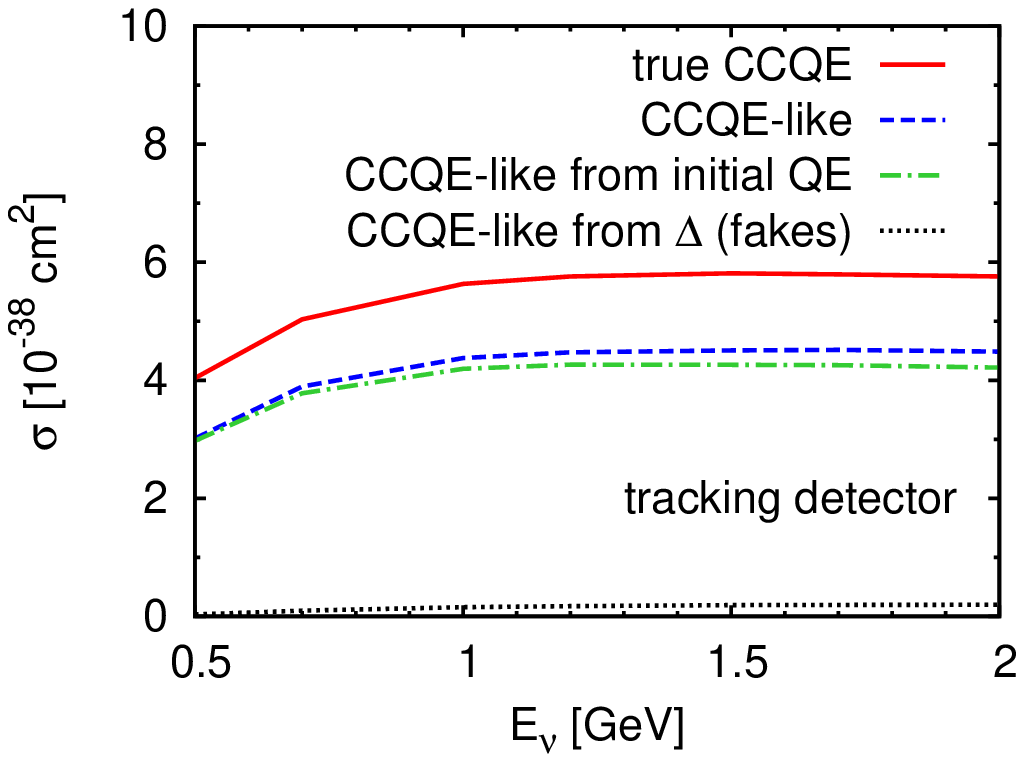}
  \caption{(Color online) Total QE cross section on \carbon{} (solid lines) compared to
    different methods on how to identify CCQE-like events in experiments (dashed lines).
    The left panel shows the method commonly applied in Cherenkov detectors; the right
    panel shows the tracking-detector method as described in the text. The contributions to the
    CCQE-like events are also classified [CCQE-like from initial QE (dash-dotted) and from
    initial $\Delta$ (dotted lines)]. Experimental detection thresholds are not taken into
    account (\cite{Leitner:2010kp}).
    \label{fig:QEmethods}}
\end{figure}

Fig.\ \ref{fig:QEmethods} shows the cross sections for scattering on $^{12}C$ as a function of neutrino energy both for the Cherenkov ($\mu,\, 0\pi$) and the tracking detector ($\mu,\, 1p, 0\pi$) identification methods \cite{Leitner:2010kp} for QE-like events. The solid curve in both cases shows the true CCQE cross section whereas the dashed line gives the QE-like cross section. It is seen that the Cherenkov detector identification method leads to a cross section that is about 20\% \emph{higher} than the true value; the surplus is due to primary resonance or pion production and subsequent fsi leading to pionless final states. For determination of the axial mass or for the reconstruction of the incoming neutrino energy this artificial surplus has to be removed with the help of an event generator. The opposite is the case for the tracking detector. Here the QE-like cross section is about 20\% \emph{lower} than the true value; this is because the secondary neutrons are not detectable. However, the QE-like sample is very 'clean' in that it contains nearly only original QE events.

\begin{figure}[tbp]
  \centering
  \includegraphics[scale=\plotscale]{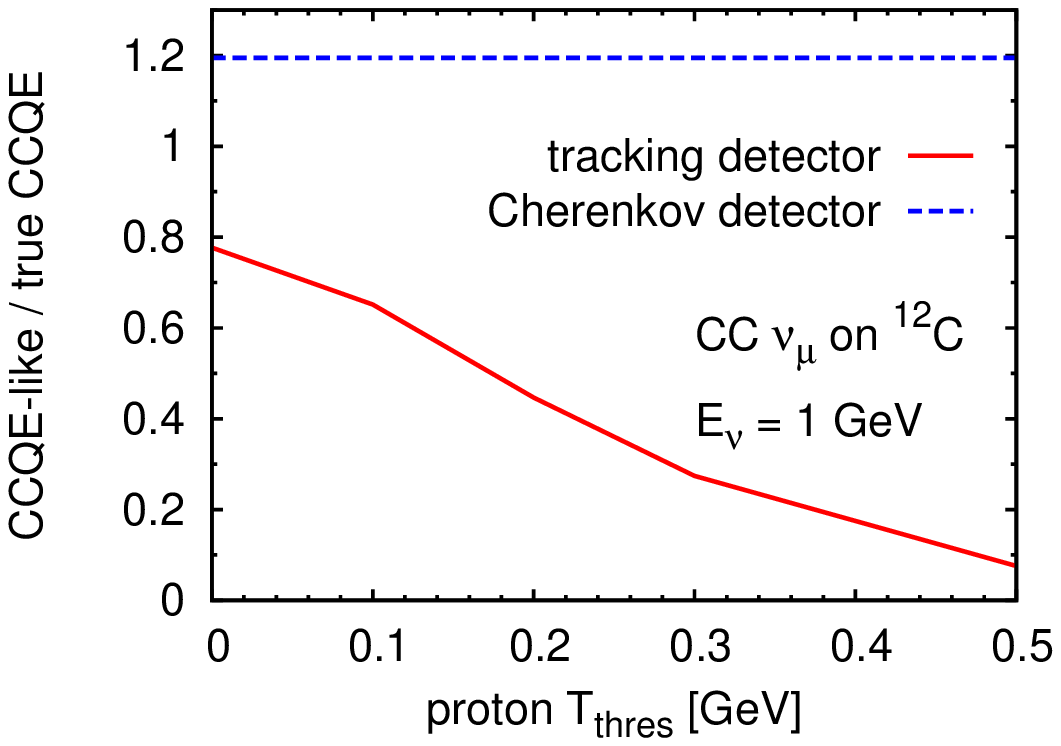}
  \includegraphics[scale=\plotscale]{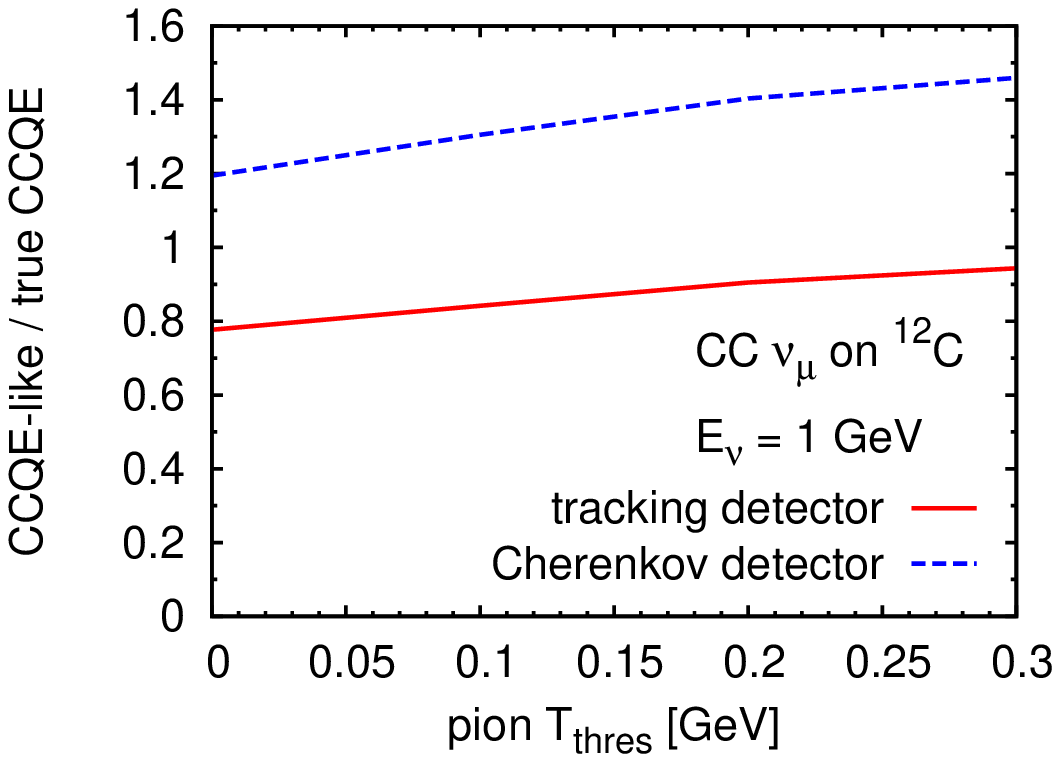}
  \caption{(Color online) Ratio of the CCQE-like to the true CCQE cross section as a function of
  the lower proton (pion) kinetic energy detection threshold for CC $\nu_\mu$ on \carbon{} at
  $E_\nu=1$ GeV. The solid lines are obtained using the tracking detector identification,
  whereas the dashed lines are for Cherenkov detectors (from \cite{Leitner:2010kp}).
  \label{fig:CCQElike_over_trueCCQE_thresholds}}
\end{figure}
A further complication arises because detectors are not perfect but have acceptance thresholds. In Fig.\ \ref{fig:CCQElike_over_trueCCQE_thresholds} we show the ratio of QE-like to true QE events as a function of proton kinetic energy treshold $T$. It is seen that for a Cherenkov detector this ratio of $\approx 1.2$ is rather independent of the proton detection threshold. On the contrary, for the tracking detector this ratio is a steeply falling function of $T$ while the ratio depends much less on the pion detection threshold (right part of Fig.\ \ref{fig:CCQElike_over_trueCCQE_thresholds}). For the thresholds of about 175 MeV for the SciFi detector and about 75 MeV for the T2K tracking detector the ratios are about 50\% or 70\%, respectively. This means that a large part of the cross section is missing and has to be reconstructed by means of a generator, making the physics result depend to a large part on the quality of this generator. A similar result holds for pion production cross sections (for a detailed discussion see \cite{Leitner:2010kp}).
\renewcommand{\plotscale}{.70}
\begin{figure}[hbp]
  \centering
  \includegraphics[scale=\plotscale]{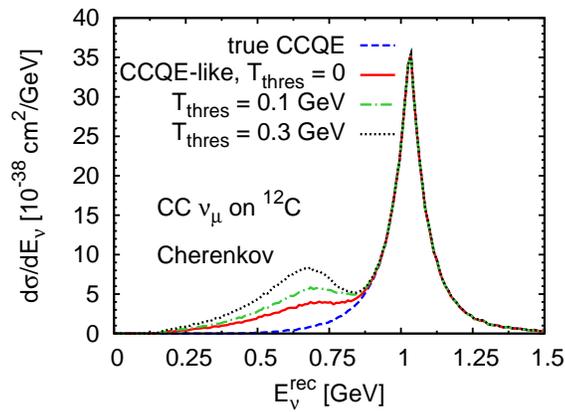}
  \caption{(Color online) Distribution of the reconstructed neutrino energy using quasifree kinematics
    on a nucleon at rest  for $E_\nu^\text{real}=1$ GeV. Using only true CCQE events for the
  reconstruction leads to the dashed line. Including CCQE-like events (Cherenkov
  definition) with various charged pion detection thresholds, one obtains the solid line
  (no pion threshold), the dash-dotted line (100 MeV), and the dotted line (300 MeV) (from \cite{Leitner:2010kp}).
  \label{fig:CC_recEnu_Cherenkov_pionThres}}
\end{figure}

We now discuss the quality of the energy reconstruction which is based on applying quasifree kinematics of true QE-scattering to QE-like events and neglecting Fermi motion \cite{Abe:2011sj}. Fig.\ \ref{fig:CC_recEnu_Cherenkov_pionThres} shows the distribution of reconstructed energies, obtained in a GiBUU simulation, for a fixed incoming neutrino energy of 1 GeV. The distribution is clearly affected by the two effects discussed above: Fermi motion leads to a broadening of the neutrino energy around the incoming energy; this accounts for the major peak at 1 GeV. The dashed curve gives the distribution of true QE events. Its width of about 16\% is determined by Fermi motion alone and is thus always present; this defines the lower limit for any energy reconstruction via QE scattering. In addition the reconstructed energy distribution exhibits a clear bump at lower energies; this is caused by an initial pion production. The bump does depend on the pion detection threshold and amounts to about 15\% of the true QE peak height at a realistic detection treshold of about 100 MeV pion kinetic energy. Taking this lower-energy bump into account raises the rms energy-width to about 22 \% (for a more detailed discussion see \cite{Leitner:2010kp}). A qualitatively similar result has also been obtained by Tanaka (see Fig.\ 3 in \cite{Tanaka:2009zzb}) using the GENIE event generator; there the lower-energy bump is even more pronounced.
\begin{figure}[th]
  \centering
  \includegraphics[scale=\plotscale]{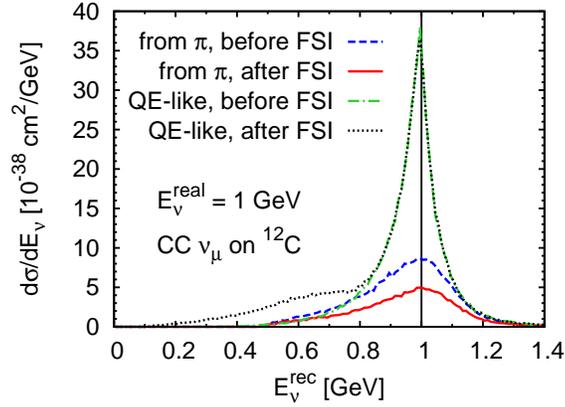}
  \caption{(Color online) Distribution of the reconstructed neutrino energy using quasifree kinematics
    on a nucleon at rest for QE and for static $\Delta$ excitation for $E_\nu^\text{real}=1$ GeV. Shown is the reconstruction based on the CCQE-like sample (before and after fsi and Cherenkov assumptions) and based
    on the CC$1\pi^+$ sample (before and after fsi). Notice that the dash-dotted and dotted curves partially overlap (from \cite{Leitner:2010kp}).
    \label{fig:energyrec_ratio}}
\end{figure}

The distribution of reconstructed neutrino energies for pion production events leads again to a rms energy-width of about 20\% \cite{Leitner:2010kp} (see Fig.\ \ref{fig:energyrec_ratio}). Since this uncertainty is unrelated to that inherent in the energy reconstruction for QE events, the overall uncertainty for the often-plotted ratio $1 \pi/CCQE$ as a function of neutrino energy also has an inherent inaccuracy of about $20 \% \times \sqrt{2} = 30 \%$ in its energy axis. Taking this properly into account can tilt the ratio as a function of neutrino energy quite significantly.
\begin{figure}[htb]
  \includegraphics[scale=\plotscale]{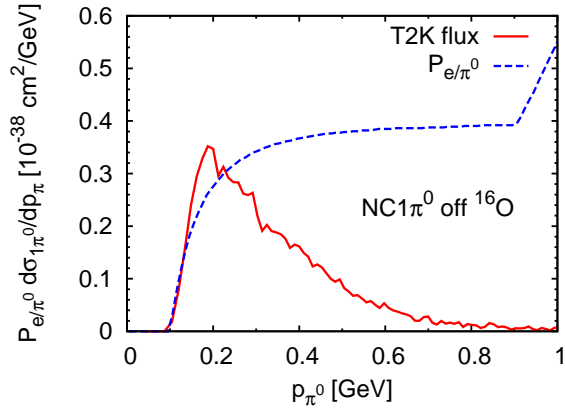}
  \caption{(Color online) NC induced single-$\pi^0$ production cross sections on \oxygen{}
    averaged over the T2K flux and multiplied by the misidentification
    probability (dashed line, taken from \refcite{Okamura:2010zz}) as a function of the
    pion momentum (from \cite{Leitner:2010jv}).
    \label{fig:T2K_NC_misid}}
\end{figure}
We have just seen that the reconstructed energies carry an inherent error of \emph{at least} about 16\%, both for QE and for pion production.

These estimates rely on the impulse approximation. It has, however, recently been shown that the impulse approximation (IA) does not describe the full total neutrino cross section in the quasielastic region and that the identification method of the MiniBooNE experiment allows for a significant amount of non-QE $2p-2h$ excitations in the QE-like cross section \cite{Martini:2009uj,Nieves:2011pp,Nieves:2011yp} that had not been removed by an event generator. These effects are not incorporated in the results above. Since for the $2p-2h$ excitations the quasifree kinematics formulas used by experiment for the energy reconstruction do not apply, the actual energy uncertainty may even be larger. This point has not yet been studied in any detail.

The T2K experiment has recently reported the first experimental observation of electron neutrino appearance from a muon neutrino beam \cite{Abe:2011sj}. At the flux maximum the energy of is $\approx 600$ MeV. At this energy the error in the reconstructed energy, according to Table I in \cite{Leitner:2010kp}, amounts to about 21\% for the Cherenkov detector. Fig.\ \ref{fig:QEmethods} shows that at 600 MeV the QE events are fairly clean, but that the admixture of pion production events rises significantly, already up to 1 GeV, reflecting the pion production threshold.

A crucial problem in any such experiment with Cherenkov counters is that the decay photons
of neutral pions can be misidentified as electrons. We therefore, investigate in
\reffig{fig:T2K_NC_misid} the probability that a misidentified $\pi^0$ is counted as a
$\nu_e$ appearance event.  The probability that a $\pi^0$ cannot be distinguished from
$e^\pm$ is given by the dashed line (taken from Fig.~2 of \refcite{Okamura:2010zz}). The solid line shows the
weighted cross section averaged over the T2K flux and calculated using the elementary ANL
data as input. The total cross section for misidentified events is now
$0.09\,\cdot\,10^{-38}\, \text{cm}^2$ and thus 26\% of the true pion events. Again, this number does not include any $2p-2h$ excitations.

\section{Summary}
Broad-band long-baseline experiments with neutrino beams necessarily average over many different reaction types. This places particular requirements on any reliable event generator. Using GiBUU we have discussed the intimate entanglement of quasielastic scattering on target nucleons and of resonance excitations, which have to be disentangled for a determination of the neutrino energy. We have discussed the achievable accuracies for the latter. Fermi-motion alone already gives a minimum energy width of about 16\% (at 1 GeV), independent of any special generator used. Misidentification of events raises this number up to 20 - 25\%, larger than previously discussed values. Furthermore, in tracking detectors the sensitivity of QE identification to experimental detection thresholds is found to be large. The contribution of NC $\pi^0$ production to misidentification of electrons in T2K is found to amount to 26\% of the total pion events. All of these numbers do not take into account any $2p-2h$ effects and are thus lower limits.

\section{Acknowledgment}
The authors gratefully acknowledge helpful comments on the T2K experiment by K.~McFarland, P.~Paul, F.~Sanchez and C.~Walter.
This work was supported by Deutsche Forschungsgemeinschaft and HIC for FAIR.

\end{document}